\documentclass[twocolumn,aps,showpacs,preprintnumbers,
superscriptaddress,
nofootinbib,
]{revtex4-2}

\usepackage{graphicx}
\usepackage{dcolumn}
\usepackage{bm}
\usepackage[utf8]{inputenc}
\usepackage{amsmath,amssymb}
\usepackage{graphicx}
\usepackage{xcolor}
\usepackage{verbatim}
\usepackage{lipsum}
\usepackage{soul}
\usepackage{hyperref}
\usepackage[mathlines]{lineno}
\usepackage[compat=1.1.0]{tikz-feynman}

\hypersetup{
     colorlinks   = true,
     citecolor    = red,
     urlcolor     = red,
     linkcolor    = red
}

\usepackage{listings}
\usepackage{color,xcolor}

\newcommand{\nn}{\nonumber \\}

\usepackage[normalem]{ulem}

\begin{document}
\preprint{PITT-PACC-2413}

\title{Sub-MHz Radio Background from Ultralight Dark Photon Dark Matter}
\author{Javier F. Acevedo}
 \thanks{ \href{mailto:jfacev@stanford.edu}{jfacev@stanford.edu}; \href{https://orcid.org/0000-0003-3666-0951}{0000-0003-3666-0951}}
\affiliation{Particle Theory Group, SLAC National Accelerator Laboratory, Stanford, CA 94035, USA}
\author{Amit Bhoonah}
\thanks{ \href{mailto:amit.bhoonah@pitt.edu}{amit.bhoonah@pitt.edu}; \href{https://orcid.org/0000-0002-4206-215X}{0000-0002-4206-215X}}
\affiliation{Department of Physics, University of Pittsburgh, Pittsburgh, PA 15260, USA}
\author{Kun Cheng}
 \thanks{ \href{mailto:kun.cheng@pitt.edu}{kun.cheng@pitt.edu}; \href{https://orcid.org/0000-0002-4959-2997}{0000-0002-4959-2997}}
\affiliation{Department of Physics, University of Pittsburgh, Pittsburgh, PA 15260, USA}

\begin{abstract}
Dark photons are a well-motivated candidate for dark matter, but their detection becomes challenging for ultralight masses with both experimental and astrophysical probes. In this work, we propose a new approach to explore this regime through the dark inverse Compton scattering of ultralight dark photons with cosmic ray electrons and positrons. We show this process generates a potentially observable background radiation that is most prominent at frequencies below MHz. We compute this effect using the latest cosmic ray models and radio absorption maps. Comparing it to observations of the Milky Way’s radio spectrum from Explorer 43, Radio Astronomy Explorer 2, and the Parker Solar Probe, we place leading constraints on the kinetic mixing of dark photon dark matter for masses $\lesssim 2 \times 10^{-17} \ \rm eV$. 
\end{abstract}
\maketitle

\section{Introduction}
\label{sec:Introduction}
While abundant evidence points to the dominant form of matter in the universe being non-luminous, little is known about the particle nature of this dark matter. Well-motivated candidates such as Weakly Interacting Massive Particles (WIMPs) or QCD axions \cite{Kim:1979if,Shifman:1979if,Dine:1981rt,Zhitnitsky:1980tq} continue to be searched for, but it remains important to explore other viable models. One such example, originally inspired by large volume string compactifications \cite{Burgess:2008ri,Goodsell:2009xc}, arises in
simple extensions of the Standard Model (SM) gauge group by an extra local $U(1)$ symmetry. The model has, in addition to the SM Hypercharge, a \emph{massive} Abelian gauge boson $A^{\prime}$. The pair mix kinetically with a small, often loop induced \cite{Pospelov:2008jk,Arkani-Hamed:2008hhe}, mixing parameter $\epsilon$. For dynamical processes occurring well below the $Z$ boson mass, the interaction Lagrangian, after diagonalization, is \cite{Dubovsky:2015cca} 
\begin{widetext}
\begin{equation}
\mathcal{L} = \mathcal{L}_{\rm SM}- \frac{1}{4}F_{\mu\nu}F^{\mu\nu} -\frac{1}{4}F^{\prime}_{\mu\nu}F^{\prime\mu\nu} + \frac{1}{2}m_{A^\prime}^{2}A^{\prime}_{\mu}A^{\prime\mu} - \frac{e}{\sqrt{1+\epsilon^{2}}}\left(A_{\mu} + \epsilon A^{\prime}_{\mu}\right)J_{\rm EM}^{\mu}~.
\end{equation}  
\end{widetext}
When this \emph{dark photon} has a mass well below that of the electron, $m_{A^{\prime}} \ll 2m_{e}$, its only decay mode is to three photons, a process with a long lifetime stable on the order of the age the universe \cite{Pospelov:2008jk}. This makes light dark photons a plausible candidate for dark matter, and various mechanisms to produce the correct relic abundance have been proposed \cite{Nelson:2011sf,Arias:2012az,Graham:2015rva,Agrawal:2018vin,Co:2018lka,Dror:2018pdh,Long:2019lwl,Capanelli:2024rlk}. 

When $m_{A^{\prime}} \ll 1$ eV, \emph{dark photon dark matter} (DPDM) is of a qualitatively different nature compared to WIMP or WIMP-like dark matter. Its number density is so large that it behaves as a quasi-coherent classical field instead of a collisionless gas of massive photons. In this regime, laboratory experiments searching for weak, dark photon-sourced electric or magnetic fields face significant challenges, since any conducting wall used to shield unwanted background electromagnetic fields also suppresses the signal field \cite{Graham:2014sha}. These effects are particularly pronounced for $m_{A^{\prime}} \lesssim 10^{-7} \ \text{eV}$, where probes like haloscopes \cite{Arias:2012az,Ghosh:2021ard} and fifth force searches \cite{Kroff:2020zhp,Bartlett:1988yy,Williams:1971ms,Tu:2005ge} lose sensitivity. On the other hand, the kinematics of ultralight dark photon interactions with astrophysical or cosmological systems generally imply the energy of the final state produced will be tiny, making it challenging to identify a signature with discovery potential. 

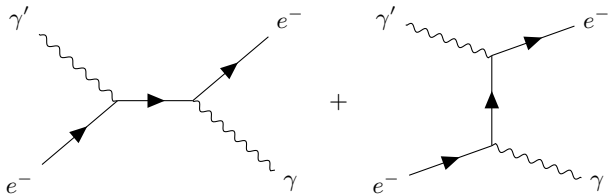
\begin{figure}[t!]
    \centering
        \begin{tikzpicture}[baseline=(center.base)]
        \begin{feynman}
            \vertex (i1) at (-1.8cm,0pt){$\gamma'$};
            \vertex (i2) at (-1.8cm,-2.2cm){$e^-$};
            \vertex (v1) at (-0.5cm,-1.1cm);
            \vertex (v2) at (0.5cm,-1.1cm);
            \vertex (o1) at (1.8cm,0pt){$e^-$};
            \vertex (o2) at (1.8cm,-2.2cm){$\gamma$};
            \vertex (label) at (0,-3cm){};
            \vertex (center) at (0,-1.2cm){};
            \diagram{(i1)--[photon](v1),(i2)--[fermion](v1)--[fermion](v2)--[fermion](o1),(v2)--[boson](o2)};
        \end{feynman}
    \end{tikzpicture}
    ~+~
    \begin{tikzpicture}[baseline=(center.base)]
        \begin{feynman}
            \vertex (i1) at (-1.4cm,0pt){$\gamma'$};
            \vertex (i2) at (-1.4cm,-2.2cm){$e^-$};
            \vertex (v1) at (0cm,-0.5cm);
            \vertex (v2) at (0cm,-1.7cm);
            \vertex (o1) at (1.4cm,0pt){$e^-$};
            \vertex (o2) at (1.4cm,-2.2cm){$\gamma$};
            \vertex (label) at (0,-3cm){};
            \vertex (center) at (0,-1.2cm){};
            \diagram{(i1)--[photon](v1),(i2)--[fermion](v2),(o2)--[photon](v2)--[fermion](v1)--[fermion](o1)};
        \end{feynman}
    \end{tikzpicture}
    \caption{Dark inverse Compton scattering by a cosmic-ray electron, $e^- \gamma' \to e^- \gamma$ (same diagrams and cross-section for positrons).}
    \label{fig:DPCRScatt}
\end{figure}

It is desirable, therefore, to investigate scenarios where the energy of the final states is boosted by the interactions of highly relativistic visible matter with DPDM. 
Cosmic rays, whose energies tend to be much larger than the keV $-$ 100 MeV energies generally associated with astrophysical systems, can provide such a boost. This was, for instance, exploited in Ref.~\cite{Su:2021jvk}, which analyzed diffuse X-ray emission resulting from the scattering of ultra high-energy cosmic ray protons against DPDM.
In this work, we analyze the interactions of relativistic cosmic ray electrons and positrons with ultralight DPDM which, at the expense of producing final photon states of lower energy, have the distinct advantage of larger flux and interaction cross-section compared to protons. We point out that the dark inverse Compton scattering process depicted in Figure~\ref{fig:DPCRScatt} results in a diffuse, potentially detectable background of radio photons. The average photon frequency produced from the dark inverse Compton scattering is
\begin{equation}\label{eq:omegaout}
    \nu \simeq \frac{\gamma_e^2 m_{A'}}{h} = 0.93 \ {\rm MHz}\left(\frac{E_{e}}{10 \ \rm{GeV}}\right)^{2} \left( \frac{m_{A^{\prime}}}{10^{-17} \ \text{eV}}\right)~,
\end{equation}
where $h$ is the Planck constant, $E_e$ is the energy of the cosmic ray electron and $\gamma_e=E_e/m_e$ is its boost factor. We compute this effect for the first time using the most recent cosmic ray models and radio absorption maps. Utilizing existing observations of the Galactic radio spectrum performed by NASA's Explorer 43 (also called the Interplanetary Monitoring Platform, IMP-6) \cite{1973ApJ...180..359B}, Radio Astronomy Explorer 2 (RAE 2) \cite{1978ApJ...221..114N}, and more recently the Parker Solar Probe (PSP) \cite{2022A&A...668A.127P}, we set leading constraints on the kinetic mixing parameter in this regime. Furthermore, we argue for the potential use of future space-based sub-MHz radio facilities in searching for this and other ultralight dark matter candidates.
Below, we work in natural units whereby $\hbar = c = 1$.

\begin{figure*}
    \centering
    \includegraphics{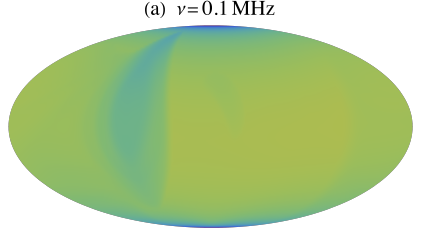}
    \includegraphics{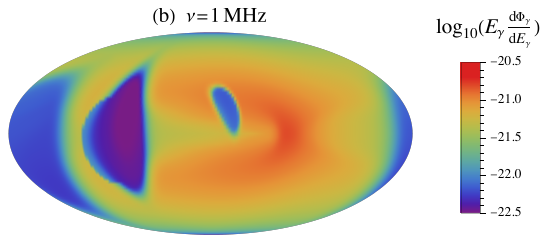}
    \caption{Mollweide projections of the flux density produced from dark inverse Compton scattering of DPDM against cosmic ray electrons for $m_{A^\prime}=10^{-19}$ eV and $\varepsilon=10^{-5}$ for the two frequencies specified, in units of $\rm W \, m^{-2} \, Hz^{-1} \, sr^{-1}$.}
    \label{fig:skymap}
\end{figure*}

\section{Dark Inverse Compton Scattering}
For the dark inverse Compton process depicted in Fig.~\ref{fig:DPCRScatt}, in the limit $m_{A'}, E_{\gamma} \ll m_e \ll E_e $, the minimum cosmic ray energy required to produce an outgoing photon of energy $E_\gamma$ is
\begin{equation}\label{eq:EcrMin}
    E_{e}^{\rm min}=m_e \, \sqrt{\frac{E_\gamma}{2m_{A'}}}.
\end{equation}
Above this threshold, in the dark photon rest frame, the differential cross-section with respect to the outgoing photon energy $E_\gamma$ is approximately
\begin{equation}
    \frac{d\sigma_{\rm DC}}{dE_\gamma} \simeq \frac{\varepsilon^2 e^4 
    }{12\pi  m_{A'}p_e^2} \left[1 - 2 \left(\frac{E_{\gamma}}{E_{\gamma}^{\rm max}}\right) 
    +2\left( \frac{E_{\gamma}}{E_{\gamma}^{\rm max}}\right)^2 
    \right], 
    \label{eq:cs_simplified}
\end{equation} 
where $p_e$ is the incoming cosmic ray electron momentum, $\varepsilon^{2} = {\epsilon^2}/{(1+\epsilon^2)}$, and $E_{\gamma}^{\rm max}$ is the maximum outgoing photon energy allowed by the kinematics of the $2\to 2$ process, 
\begin{equation}\label{eq:maxRadioWRTboost}
    E_{\gamma}^{\rm max} = 2 \gamma_e^2 m_{A'}~.
\end{equation}
The derivation of the cross-section above is included in Appendix \ref{app:ampcalc}. The associated stopping power of the DPDM medium for cosmic ray electrons is analyzed in Appendix~\ref{app:CREloss}, but we remark that it is small relative to the typical cosmic ray energies. It should be noted that other works have considered the non-relativistic limit of this process. For instance, Ref.~\cite{Hochberg:2021zrf} considered light dark matter absorption at direct detection experiments to probe much heavier DPDM. Ref.~\cite{Arza:2024iuv} analyzed the same process as here for free non-relativistic electrons in the Milky Way's interstellar medium. However, because we consider cosmic ray electrons which, in comparison, are significantly more energetic, we are able to obtain a prospective signal for much smaller dark photon masses. 

\section{Sub-MHz Radio from Ultralight Dark Photons}

For a pointwise isotropic cosmic ray flux~\cite{Fermi-LAT:2010wgm,Adriani:2015kfa}, the flux density produced by dark inverse Compton scattering within a field of view $\Delta \Omega$ is 
\begin{widetext}
\begin{eqnarray}\label{eq:fluxFullInetegral}
    E_{\gamma}\frac{d\Phi_{\gamma}}{dE_\gamma} =  \frac{1}{4\pi} \int_0^{\infty} ds \int_{\Delta \Omega} d\Omega \left(\frac{\rho_{A^\prime}(s,\Omega)}{m_{A'}}\right) \times \left( \sum_{~k=e^\pm} \int^{\infty}_{E_{e}^{\rm min}} \frac{d\Phi_{k}(s,\Omega)}{dE_{k}} \times E_{\gamma} \frac{d\sigma_{\rm DC}}{dE_\gamma} \, dE_{k} \right) \times P_{\nu}(s,\Omega) ~,
\end{eqnarray}
\end{widetext}
where $s$ is the line of sight distance, and the differential solid angle of the field of view is $d\Omega = \cos b \ db \ dl$, where $(l,b)$ are the Galactic longitude and latitude. In terms of these variables, the Galactocentric distance $R$ is $R^2 = R^2_{\rm GC}+s^2 - 2 \, s \, R_{\rm GC} \cos(b) \cos(l)$, where we take $R_{\rm GC} \simeq 8.5 \ \rm kpc$ as the distance from the Earth to the Galactic Center (our results are not sensitive to this input). The factor $\rho_{A^\prime}$ is the DPDM mass density. The innermost integral is the photon production rate due to the dark inverse Compton scattering. This is expressed in terms of the differential cosmic ray electron/positron spectrum $d\Phi_{k}/dE_{k}$ ($k = e^\pm$), and the differential cross-section to produce a final photon of energy $E_\gamma$. We sum over the primary and secondary electron fluxes, as well as primary positron flux. Finally, the factor $P_{\nu}$ incorporates absorption effects as the low-energy photons propagate through the Milky Way's interstellar medium. All of these inputs are fully detailed below. We integrate Eq.~\eqref{eq:fluxFullInetegral} using the numerical package Vegas~\cite{PETERLEPAGE1978192}. 

In terms of the optical depth $\tau_\nu$, the absorption factor is \cite{2021ApJ...914..128C}
\begin{equation}
    P_{\nu}(s,\Omega) = \exp\left[-\tau_\nu(s,\Omega)\right]~,
    \label{eq:Pnu}
\end{equation}
\begin{align}
    \tau_\nu(s,\Omega) \simeq 0.65 \, & \left(\frac{\nu}{\rm MHz}\right)^{-2.1} \left(\frac{T}{10^4 \ \rm K}\right)^{-1.35} \\ & \times \int_{0}^{s} \left(\frac{n_e(s^{\prime},\Omega)}{\rm cm^{-3}}\right)^2 \left(\frac{ds^{\prime}}{\rm pc}\right)~. \nonumber 
\end{align}
The optical depth has a strong dependence on the free electron density along the line of sight chosen. To compute its effect, we assume a typical warm interstellar medium temperature of $T \simeq 10^4 \ \rm K$, and use the global Milky Way electron density model of Ref.~\cite{2017ApJ...835...29Y}, hereafter YMW16, which accounts for thin and thick disk contributions, the spiral arms, as well as a number of observed nearby features. For our main estimates, we use the best-fit values for the various parameters of YMW16. However, we comment on the uncertainties associated with this model.  
At the solar system position, the free electron distribution is characterized by an approximately constant density region, surrounded by an overdensity structure at $\mathcal{O}(100 \ \rm pc)$ distance. 
Because our bounds are driven by low frequency observations in the range $0.1 - 0.5 \ \rm MHz$, at which the galaxy is only transparent from the observation point to $\mathcal{O}(100 \ \rm pc)$, the main source of uncertainty is the background electron density predicted by YMW16 within the Local Bubble. This can be as high as 50$\%$, leading to a factor of order two uncertainty in the radio wave flux.

For the cosmic ray spectra, we use model $^{S}S Z_{4}R_{20}T_{150}C_{2}$ from ~\cite{Fermi-LAT:2012edv}, sourced using the GALPROP software \cite{2011CoPhC.182.1156V}, which (among other models) has been found consistent with the diffuse gamma-ray emission observations of the Fermi-LAT telescope \cite{Fermi-LAT:2009ico}, as well as the electron and/or positron fluxes measured by the AMS-02 experiment \cite{AMS:2019iwo} and Voyager Spacecrafts \cite{Cummings:2016pdr,2019NatAs...3.1013S}. This model assumes the Milky Way as a cylindrical galaxy of radius 20 kpc ($R_{20}$) and height 8 kpc ($Z_{4}$), with a hydrogen spin temperature of 150 K ($T_{150}$). The remaining model parameters are the supernova remnant distribution from Ref.~\cite{1998ApJ...504..761C} ($^{S}S$) (assumed to be the main source of primary cosmic-rays), and the E(B-V) magnitude cuts used in processing galactic matter maps ($C_{2}$). It should be noted here that electrons, through bremsstrahlung, contribute only a sub-dominant fraction of the diffuse gamma ray emission, the dominant one being the production of neutral pions by cosmic ray nuclei impacting on interstellar matter and their subsequent decay to two photons. One may therefore worry that the diffuse gamma ray emission map may not accurately predict the electron and positron fluxes. We note, however, that the electron and positron fluxes adopted in Ref.~\cite{Fermi-LAT:2012edv} are consistent with those obtained using more direct tracers like diffuse synchrotron emission~\cite{Strong:2011wd}.
We use the outputs for the primary electron and positron spectra, and the one for secondary electrons produced as a result of the interaction of cosmic ray protons and helium nuclei with interstellar gas, the contribution of heavier nuclei being negligible.
The variation of the cosmic ray flux from model dependence is further detailed in Appendix~\ref{app:CRemission}. However, as before, due to the large optical depth of the Milky Way at sub-MHz frequencies only emission within $\mathcal{O}(100 \,\rm pc)$ distance would be observed, a region where the various models differ only at the percent level.

Our predicted flux is insensitive to the choice of dark matter density profile $\rho_{A^\prime}$ at sub-MHz frequencies. This is again because of the extreme optical depth of the interstellar medium, which implies at most only the first $\mathcal{O}(100 \,\rm pc)$ along the line of sight effectively contribute to the flux. Within such small distance, the choice of profile only impacts our estimates at the percent level. For concreteness, we have considered an Einasto profile \cite{1965TrAlm...5...87E}, with slope $\alpha=0.17$ and scale radius $R_s=20$~kpc, based on the Milky Way-like halos in the DM-only Aquarius simulations~\cite{Pieri:2009je}. At our local position, we fix $\rho_{A^\prime}(R = R_{\rm GC}) = 0.42 \ \rm GeV \ cm^{-3}$ \cite{necib,eilers2019} (we assume dark photons are all of the dark matter, see below if this assumption is relaxed).

Figure~\ref{fig:skymap} shows the morphology of the sub-MHz radio background sourced by cosmic ray-DPDM scattering across the sky for two different frequencies, for a benchmark dark photon mass and kinetic mixing value. At 0.1 MHz, the interstellar medium's extreme opacity renders the signal weaker but nearly isotropic across the sky, except at high Galactic latitudes, where the cosmic ray flux rapidly decreases, and a nearby highly opaque local region at longitudes $l \sim -90^{\circ}$. At 1 MHz, the flux becomes highly anisotropic, with a number of discernible features associated with the inhomogeneous electron density structure within $\sim \rm kpc$ distance. Notably, regions of higher opacity near the Galactic Center direction correspond to local electron overdensities that obscure much of the emission behind them. The Carina-Sagittarius arm emerges between these structures. Towards the anti-center direction, a large electron overdensity beyond the Local Bubble depression obscures most of the emission. As with lower frequencies, the Galactic Poles remain the least opaque, though the emission drops due to the reduced cosmic ray flux at high latitudes. Future work could explore using this frequency-dependent anisotropy to boost sensitivity.

\begin{figure}[t!]
    \centering
    \includegraphics[width=0.99\linewidth]{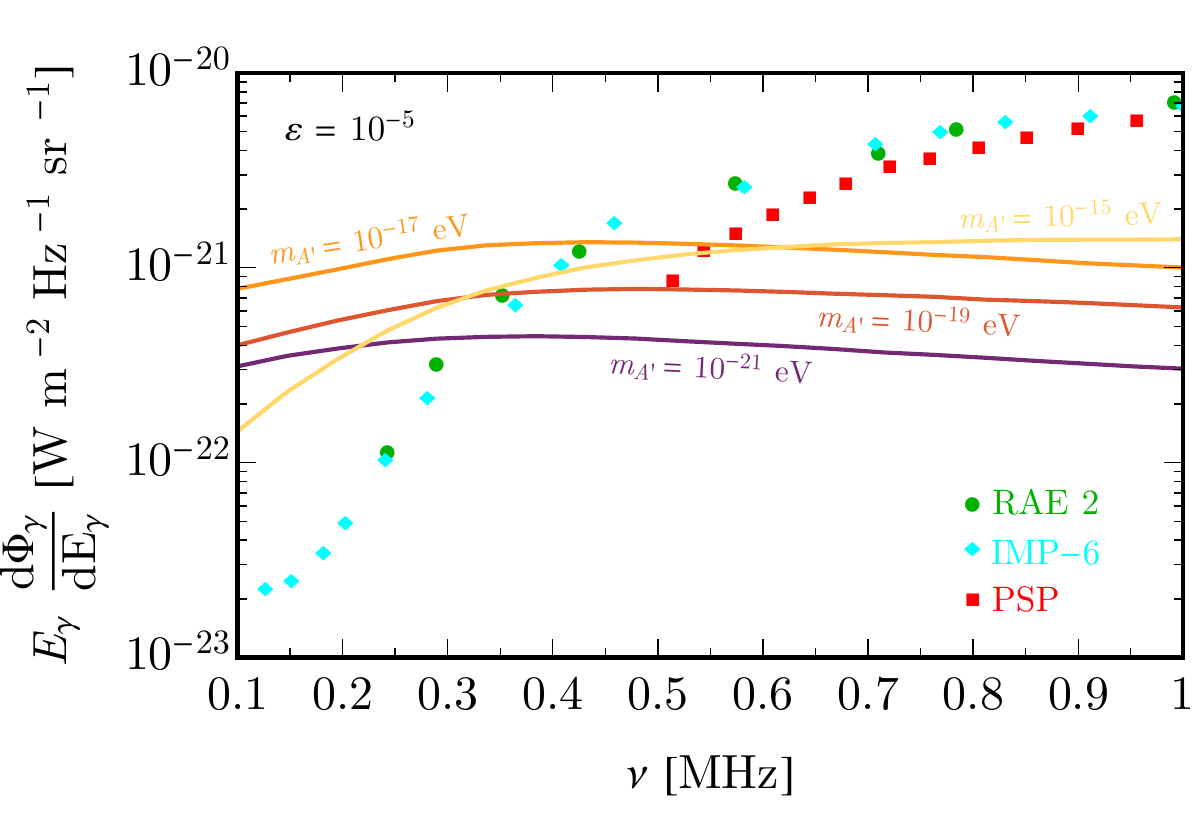}
    \caption{Mean sky brightness from cosmic ray electron$-$DPDM inverse Compton scattering, for various dark photon masses and kinetic mixing parameter as specified. Overlapped are mean sky Galactic radio spectrum observations performed by IMP-6, RAE 2, and PSP in the 0.1 $-$ 1 MHz frequency range.}
    \label{fig:DP_flux}
\end{figure}

Figure~\ref{fig:DP_flux} shows the sky-averaged brightness, obtained from integrating Eq.~\eqref{eq:fluxFullInetegral} over all directions and dividing by the spanned field of view, for a fixed kinetic mixing parameter and a range of dark photon masses. For comparison, we have included the Milky Way's radio spectrum measurements from IMP-6, RAE 2 and PSP. For each mass value, the different features can be understood from the interplay between dark photon-cosmic ray kinematics and the cosmic ray flux.
For a dark photon mass around $10^{-21}$ eV, by Eq.~\eqref{eq:omegaout}, the required cosmic ray energy for radio frequencies between $0.1 - 1$ MHz is around $0.3 - 1$ TeV. This energy range coincides with the high-energy tail of the electron cosmic ray flux, where it rapidly decreases with energy, and so the radio brightness slightly decreases with increasing frequency. By contrast, for a dark photon mass around $10^{-15}$ eV, the cosmic ray energy required is about $0.3 - 1$ GeV to produce the same radio frequency, corresponding instead to the low-energy tail of the cosmic ray spectrum, which rises steeply with energy. In this case, the radio brightness increases with frequency. For dark photon masses ranging $10^{-17}-10^{-19} \ \rm eV$, an intermediate regime is obtained. 

\begin{figure}
    \centering
    \includegraphics[width=0.99\linewidth]{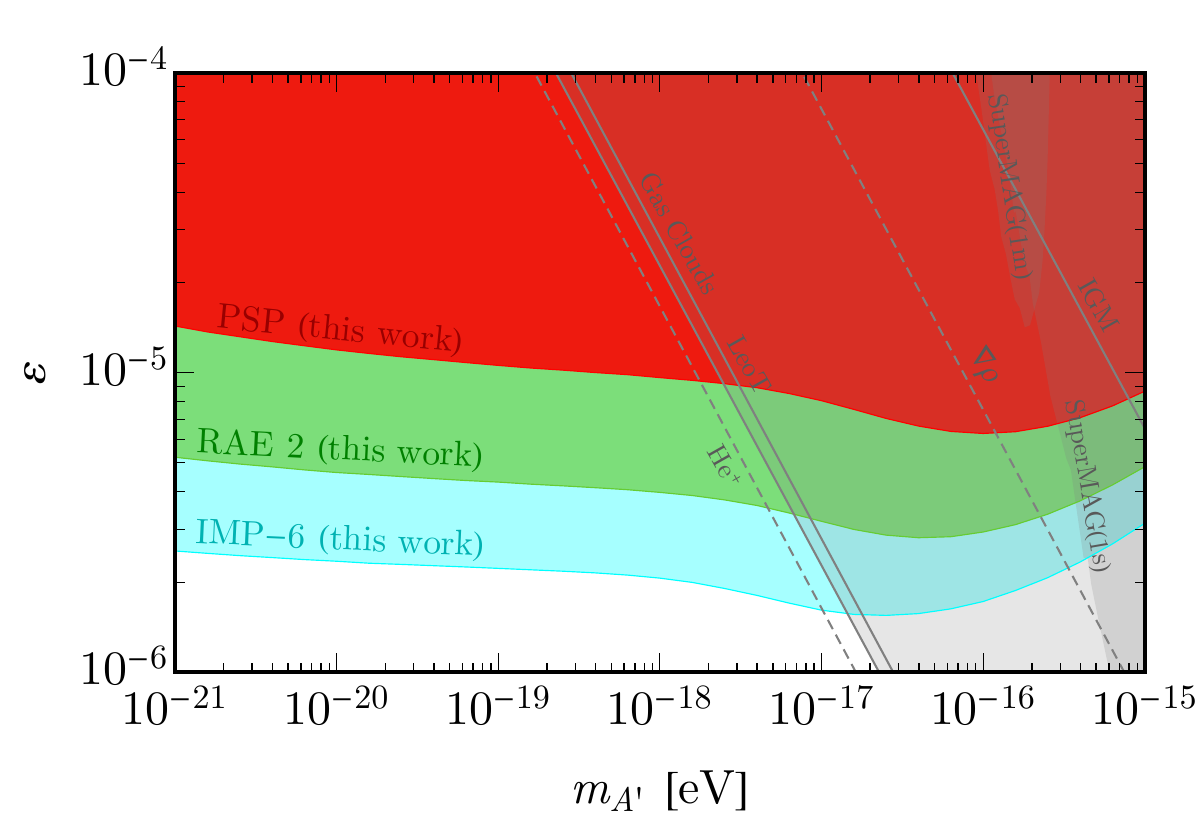}
    \caption{Constraints on the kinetic mixing parameter as a function of dark photon mass derived from IMP-6, RAE 2 and PSP radio observations, assuming dark matter is entirely composed of ultralight dark photons (otherwise, they scale linearly with the assumed fraction). See the text for complementary experimental, astrophysical and cosmological limits.}
    \label{fig:DP_bounds}
\end{figure}

Figure~\ref{fig:DP_bounds} shows the inferred limits on the kinetic mixing parameter using IMP-6, RAE 2, and PSP measurements, as a function of dark photon mass. These values are excluded by requiring the resulting brightness from DPDM-cosmic ray scattering not to exceed the observed brightness by any of these probes in each frequency bin (in the case of IMP-6, we demand the value not to exceed the probable maximum spectrum observed). This procedure is conservative as we have not incorporated additional astrophysical backgrounds into our analysis, which would produce more stringent limits. We terminate our bounds at $10^{-21} \ \rm eV$, since for masses below this threshold, stringent limits from Lyman-$\alpha$ measurements apply \cite{Irsic:2017yje,Nori:2018pka}. Moreover, masses in the range $10^{-21} - 10^{-19} \ \rm eV$ are disfavored by small-scale structure observations \cite{Hui:2016ltb,Schutz:2020jox,DES:2020fxi,Dalal:2022rmp}, and ultralight dark matter can only be a sub-component of the total relic abundance. For simplicity, we have plotted our constraints always under the assumption that DPDM saturates the relic abundance, but we note the emission scales linearly with the assumed fraction, $cf.$ Eq.~\eqref{eq:fluxFullInetegral}, producing strong limits even in this sub-component scenario. We also show complementary constraints from gas cloud heating \cite{Bhoonah:2018gjb}, Leo T \cite{2021PhRvD.103l3028W}, intergalactic medium heating \cite{Dubovsky:2015cca}, and the super-MAG dark photon search \cite{Fedderke:2021rrm,Friel:2024shg}. The dashed lines show cosmological constraints based on resonant conversion of dark photons in the primordial plasma \cite{Arias:2012az,McDermott:2019lch,Witte:2020rvb,Caputo:2020rnx,Caputo:2020bdy}. 

At masses $m_{A^{\prime}} \lesssim 10^{-17} \ \rm eV$, the observed sub-MHz radio spectrum imposes the strongest limit, reaching down approximately to $\varepsilon \lesssim 2\times 10^{-6}$ in the case of IMP-6. Our sensitivity decreases with increasing dark photon mass, as this implies a smaller number density of targets for the cosmic rays, and emission peaks at frequencies where the measured flux is larger. On the other hand, our sensitivity also decreases with decreasing mass. This is because the cosmic ray energy threshold to produce a photon of frequency $\gtrsim 0.1 \ \rm MHz$ is increased, implying a smaller fraction of the cosmic ray spectrum will contribute to the emission. This sensitivity loss, however, is weaker compared to the regime $m_{A^{\prime}} \gtrsim 10^{-17} \ \rm eV$, as the emission peaks at lower frequencies where the observed brightness also decreases. 
The overall uncertainty in our estimates derived from the degeneracy between possible cosmic ray models and Galactic dark matter profiles only reaches the percent level. This is largely due to absorption effects, which imply only the contribution from the first $\mathcal{O}(100 \ \rm pc)$ along the line of sight, where the uncertainties are the lowest, is relevant. As outlined above, the largest source of uncertainty comes from the free electron map parameters of YMW16, which translates into an $\mathcal{O}(1)$ variation in the limits derived.

\section{Summary and Outlook}
We have analyzed the interactions of cosmic ray electrons and positrons with ultralight dark photon dark matter, and shown that dark inverse Compton scattering results in a diffuse, almost isotropic flux of radio photons most prominent at sub-MHz frequencies. We have computed this flux using state-of-the-art cosmic ray models which reproduce a number of independent observations, as well as the latest electron density maps of the Milky Way to account for the absorption of this background by the interstellar medium. Despite the high opacity of the Milky Way to radio waves at these frequencies, we have found that this flux can be significant, especially for dark photon masses that remain challenging to probe with terrestrial experiments. Utilizing Galactic spectrum observations from NASA's Explorer 43 (IMP-6), the Radio Astronomy Explorer 2, and the Parker Solar Probe, we have set leading constraints on the kinetic mixing parameter of ultralight dark photon dark matter for masses $m_{A^\prime} \lesssim 2 \times  10^{-17} \ \rm eV$, reaching $\varepsilon \lesssim (2 - 10) \times 10^{-6}$ depending on the dataset.

Future improvements in sensitivity can be achieved by incorporating additional backgrounds in this region of the spectrum, including extragalactic sources, as well as free-free and synchrotron emissions from the interstellar and circumgalactic media.
Additionally, it will be interesting to explore the potential sensitivity of upcoming or in-development missions designed to probe this decameter wavelength regime, such as the Sun Radio Interferometer Space Experiment (SunRISE) mission \cite{9843607}, the  Nanosatellites pour
un Observatoire Interférométrique Radio dans l’Espace (NOIRE) project \cite{2017arXiv171010245C}, or the lunar-based Large-scale Array for Radio Astronomy on the Farside (LARAF) \cite{Chen:2024tvn} and Lunar Surface Electromagnetics Experiment at Night (LuSEE-Night) \cite{Saliwanchik:2024nrp}. Beyond dark photons, other light relics such as axion-like particles and dilatons, with couplings to both photons and electrons, could produce a similar diffuse background through analogous processes with cosmic rays, if their masses are in a similar range. Moving forward, it will be important to consider how this effect can be used to search for these other models of dark matter, and to identify any features therein capable of distinguishing between them.

\begin{acknowledgments}
We thank Brian Batell, Joshua Berger, Joseph Bramante, Rebecca Leane and Tom Rizzo for helpful comments and discussions. JFA is supported in part by the U.S. Department of Energy under Contract DE-AC02-76SF00515.
KC and AB are supported in part by the U.S. Department of Energy under grant No. DE-SC0007914. AB also acknowledges support by the IQ Initiative at the University of Pittsburgh. 
\end{acknowledgments}

\clearpage
\newpage
\onecolumngrid

\appendix
\section{Dark Inverse Compton Scattering Amplitude}\label{app:ampcalc}

In the dark photon rest frame, otherwise equivalent to the lab frame given that they are non-relativistic, we write the four-momenta as
\begin{align}
    p_1 &= (E_e,0,0,p_e) \\
    k_1 &= (m_{A'},0,0,0) \\
    p_2 &= (E_{e}',0 ,p_e' \sin\theta_e, p_e'\cos\theta_e ) \\
    k_2 &= (E_{\gamma},0,E_{\gamma}\sin\theta,E_{\gamma}\cos\theta)~,
\end{align}
where the assignment is as follows: $p_{1}$ for the incoming cosmic ray four momentum (assumed to be traveling in the $\hat{z}$ direction), $k_{1}$ for that of the dark photon, and $p_{2}$ and $k_{2}$ for respectively those of the outgoing cosmic ray and photon. In the lab frame, the energy of outgoing photon is related to its scattering angle as
\begin{equation}
    E_\gamma = \frac{m_{A'}^2+2 m_{A'} E_e}{2(E_e+m_{A'}-p_e\cos\theta)}~,
\end{equation}
%
so that the kinematic limits of the photon energy are
\begin{align}
    E_\gamma^{\rm min}&=\frac{m_{A'}^2+2 m_{A'} E_e}{2(E_e+m_{A'}+p_e)} \simeq \frac{m_{A'}}{2}, \label{eq:Egam_min} \\
    E_\gamma^{\rm max}&=\frac{m_{A'}^2+2 m_{A'} E_e}{2(E_e+m_{A'}-p_e)}\simeq 2m_{A'}\left(\frac{E_e}{m_e}\right)^2. \label{eq:Egam_max}
\end{align}
The differential cross section in this frame is
\begin{equation}
   \frac{d\sigma_{\rm DC}}{dE_\gamma} = \frac{1}{32\pi m_{A'}p_e^2} \, \overline{\sum}|\mathcal{M}|^2, 
\end{equation}
with the spin-averaged matrix element square given by
\begin{align}
    \overline{\sum} |\mathcal{M}|^2 =& 
    \frac{1}{6}\sum |\mathcal{M}|^2 \nn
= & \frac{e^4 \varepsilon^2}{6  m_{A^{\prime}}^2\left(2 E_e+m_{A^{\prime}}\right)^2\left(2 E_e+m_{A^{\prime}}-2 E_\gamma\right)^2} \nn
\times & \Big[m_{A^{\prime}}^2\left(2 E_e+m_{A^{\prime}}\right)^2\left(2 E_e^2+2 E_e m_{A^{\prime}}+m_e^2+m_{A^{\prime}}^2\right) \nn
& +E_\gamma^2\left(m_{A^{\prime}}^2\left(2 E_e+m_{A^{\prime}}\right)\left(6 E_e+5 m_{A^{\prime}}\right)+4 m_e^4+2 m_e^2 m_{A^{\prime}}\left(4 E_e+3 m_{A^{\prime}}\right)\right)  -2 m_{A^{\prime}}^2 E_\gamma^3\left(2 E_e+m_{A^{\prime}}\right) \nn
& -2 m_{A^{\prime}} E_\gamma \left(2 E_e+m_{A^{\prime}}\right)\left(4 E_e^2 m_{A^{\prime}}+E_e\left(2 m_e^2+5 m_{A^{\prime}}^2\right)+2 m_{A^{\prime}}\left(m_e^2+m_{A^{\prime}}^2\right)\right)\Big]~.
\end{align}
For the range of dark photon masses, electron energies, and radio frequencies we focus on, it is always the case that
\begin{equation}
    \frac{m_{A'}}{E_e}\lesssim\frac{E_\gamma}{E_e}<\frac{E_\gamma}{m_e} < 10^{-11}~,
\end{equation}
and the maximum energy of outgoing photon with respect to the dark photon mass is enhanced by the square of the comic ray boost factor $\gamma_e = E_e/m_e$,
\begin{equation}
    E_\gamma \sim \gamma_e^2 \, m_{A'}~.
\end{equation}
Defining
\begin{equation}
    x\equiv \left(\frac{1}{\gamma_e}\right)^2 \frac{E_\gamma}{m_{A'}} = \frac{E_\gamma}{2E_\gamma^{\rm max}},
\end{equation}
which is an $\mathcal{O}(1)$ variable, we rewrite the squared amplitude as
\begin{align}
    \overline{\sum} |\mathcal{M}|^2 = \frac{2e^4 \varepsilon^2}{3} B^2 \Bigg[ &A^2 \left( 2x^2 + \frac{m_{A'}E_\gamma}{E_e^2}x\right) +A\left(\frac{2m_{A'}-4E_\gamma}{E_e}\, x+\frac{E_\gamma^3+E_\gamma(E_\gamma-m_{A'})^2}{E_e^3}\right) \nn
    +& 4+\frac{1}{\gamma_e^2}-4x + \frac{2m_{A'}-4E_\gamma}{E_e} + \frac{3E_\gamma(E_\gamma-m_{A'})}{E_e^2}
    \Bigg],
\end{align}
where
\begin{align}
    A&=\frac{2E_e}{2E_e+m_{A'}}= 1+\mathcal{O}(10^{-11}), \nn
    B&=\frac{2E_e}{2E_e+m_{A'}-2E_\gamma}= 1+\mathcal{O}(10^{-11}). \nonumber
\end{align}
Then the amplitude is expanded as
\begin{equation}
    \overline{\sum}|\mathcal{M}|^2 = \frac{2e^4 \varepsilon^2}{3}\left( 4+\frac{1}{\gamma_e^2} -4x+2x^2 \right) +\mathcal{O}(10^{-11}) 
    \simeq \frac{2e^4 \varepsilon^2}{3}\left( 4 -4x+2x^2 \right)~.
\end{equation}
In the final expression, we have neglected the $1/\gamma_e^2 \ll x$ terms as well since they contribute negligibly, only introducing corrections to the dark inverse Compton scattering cross section at $\mathcal{O}(10^{-11})$.

\section{Stopping Power of Dark Photons}
\label{app:CREloss}
We estimate the stopping power of the Galactic DPDM halo due to dark inverse Compton scattering and show that the energy loss across $\sim \rm kpc$ scales is negligible for individual cosmic ray electrons or positrons. Because $m_{A^\prime} \ll E_\gamma$ in this regime, the energy lost by the cosmic ray is approximately the energy transferred to the outgoing photon. The stopping power is then given by
\begin{equation}
    \frac{dE_e}{dR} \simeq \left(\frac{\rho_{A^\prime}(R)} {m_{A^\prime}}\right) \times \int_{E_{\gamma}^{\rm min}}^{E_{\gamma}^{\rm max}} \frac{d\sigma_{\rm DC}}{dE_\gamma} E_\gamma \, dE_\gamma~,
\end{equation}
where this is a function of Galactocentric distance $R$. The minimum and maximum final state photon energies are given by Eqs.~\eqref{eq:Egam_min} and \eqref{eq:Egam_max}, respectively. Since $E_\gamma^{\rm min} \ll E_\gamma^{\rm max}$, integrating the above yields
\begin{equation}
    \frac{dE_e}{dR} \simeq \left(\frac{\rho_{A^\prime}(R)}{m_{A^\prime}}\right) \times \left(\frac{\varepsilon^2 e^4}{12 \pi m_{A^\prime} p_e^2}\right) \times \frac{\left(E_\gamma^{\rm max}\right)^2}{3}\lesssim 0.1 \ {\rm \frac{MeV}{kpc}} \left(\frac{\varepsilon}{10^{-5}}\right)^2\left(\frac{E_e}{10 \ \rm GeV}\right)^2\left(\frac{\rho_{A^\prime}}{0.42 \ \rm GeV/cm^3}\right)~.
\end{equation}
The rightmost expression has been normalized to the local dark matter density (note that this value only changes by an $\mathcal{O}(1)$ factor over $\sim \rm kpc$ scales, except when close to the Galactic Center).
This estimate demonstrates that energy losses from this process are negligible over Galactic scales compared to the typical energies of cosmic ray electrons and positrons. Therefore, we would not expect any potentially observable features in the cosmic ray electron or positron spectra. 

\section{Cosmic Ray Models}
\label{app:CRemission}

\begin{figure}
   \centering
   \includegraphics{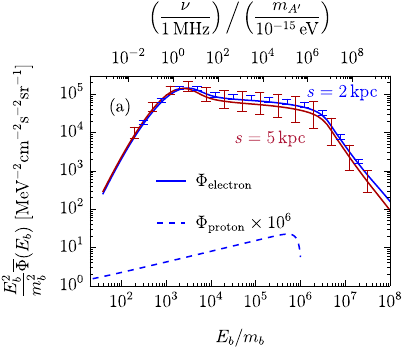}
   \includegraphics{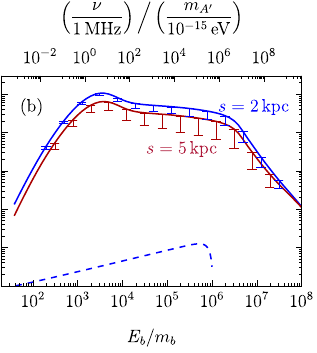}
   \includegraphics{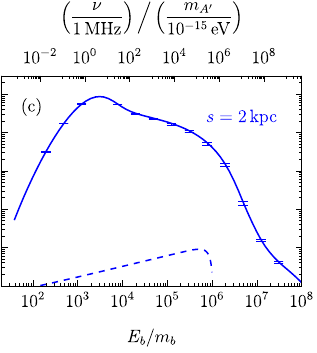}
   \caption{Integrated comic ray flux for two different line of sight distances, viewed (a) towards the galaxy center, $b=l=0^\circ$, (b) towards the anticenter, $b=0^\circ,l=180^\circ$ and (c) toward the pole, $|b|=90^\circ$. The solid line denotes the cosmic ray flux from the $^{S}S Z_{4}R_{20}T_{150}C_{2}$ model (used in main text), while the interval bands range from the minimal to maximal value calculated from the four models we considered. For comparison, we also show the integrated proton flux multiplied by a factor of 10$^{6}$, indicating that the radio flux due to dark inverse Compton scattering of protons against dark photons in the mass range considered in this work is negligible.}
   \label{fig:IntegratedCRflux}
\end{figure}

We estimate the variation in the predicted flux spanned by the various cosmic ray models considered in Ref.~\cite{Fermi-LAT:2012edv}. This can be estimated purely from the shape of the cosmic ray flux, as the amplitude square of the dark inverse Compton scattering process is an $\mathcal{O}$(1) variable in the entire energy range we consider. To this end, using Eqs.~\eqref{eq:EcrMin} and \eqref{eq:fluxFullInetegral}, we define
\begin{equation}
    \overline{\Phi}(E_{e}^{\rm min})\equiv \sum_{k = e^\pm}\int_{E_{e}^{\rm min}}^\infty \frac{d\Phi_k(s,\Omega)}{dE_k}\, \frac{(E_{e}^{\rm min})^2}{E_k^2} \,dE_k
\end{equation}
to be the integrated comic ray flux weighted by $1/E_k^2$ to account for the flux factor in the cross-section. Using this function, it becomes more convenient to evaluate the power flux $E^2_{\gamma} \frac{d\Phi_{\gamma}}{dE_{\gamma}}$, which is proportional to 
\begin{equation}
  \left(\frac{\rho_{A^\prime}(s,\Omega)}{m_{A'}}\right) \times \sum_{k = e^\pm} \int_{E_{e}^{\rm min}}^\infty \frac{d\Phi_{k}(s,\Omega)}{dE_{k}} \times E_{\gamma}^2 \frac{d\sigma_{\rm DC}}{dE_{\gamma}} \, dE_{k}   \simeq \frac{\rho_{A'}}{2\pi^2} \frac{\left(E_{e}^{\rm min}\right)^2}{m_e^4} \overline{\Phi}(E_{e}^{\rm min})~,
\end{equation}
where we treat the squared amplitude as an $\mathcal{O}(1)$ constant.
This way, the uncertainty spanned by $\bar{\Phi}(E_{e}^{\rm min})$ through the various cosmic ray models is solely dependent on the threshold $E_e^{\rm min}$, which in turn depends on the ratio $E_\gamma/m_{A^\prime}$.

Figure~\ref{fig:IntegratedCRflux} shows the resulting $\bar{\Phi}(E_{e}^{\rm min})$ for the cosmic ray models spanned by considering all four cosmic ray source distributions $^{S}S $, $^{S}L$, $^{S}Y$ and $^{S}O$ used in Ref.~\cite{Fermi-LAT:2012edv}, while assuming the geometric profile $Z_4R_{20}$ and hydrogen spin temperature $T_{150}$. These are shown for four benchmark directions: the center, anti-center and poles, at two different line of sight distances of 2 and 5 kpc. In order to show the impact cosmic ray modeling, we have chosen $^{S}S Z_{4}R_{20}T_{150}C_{2}$ as the benchmark model, and shown the maximal difference among the four models as the vertical interval bands. While the variation among models can reach an order of magnitude level at 5 kpc distance, for distances below 2 kpc they are overall negligible in all directions. Since for frequencies at MHz or below any contribution beyond kpc distances is suppressed by opacity effects, we conclude that our signal is not sensitive to the chosen cosmic ray model. We also note that $(E_{e}^{\rm min})^2 \overline{\Phi}(E_{e}^{\rm min})$ does not vary significantly with line of sight distance or field of view.

All of the above estimates are also valid for the dark inverse Compton scattering process with cosmic ray protons, as long as $m_{A'},E_{\gamma}\ll m_p$ where $m_p$ is the proton mass. However, for a cosmic ray proton to produce a radio wave of the same frequency as in the dark inverse Compton scattering process with electrons or positrons, the same boost factor is required, \textit{i.e.}, the proton cosmic ray energy should be about 2000 times larger than that of the electron. Therefore, the same process with the proton is suppressed because of the smaller flux at higher energy (the higher energy also has the compounding effect of suppressing the cross section due to the flux factor). For a more quantitative comparison, we also show the analogous function $((E_{p}^{\rm min})^2/m_p^4) \overline{\Phi}(E_{p}^{\rm min})$ for cosmic ray protons in Fig.~\ref{fig:IntegratedCRflux}, and confirm that their contribution is negligible.

\bibliography{apssamp}

\end{document}